\documentclass[12pt]{article}
\usepackage{latexsym,cite,subeqn}
\input amssym.def
\input amssym.tex

\makeatletter

\renewcommand{\thesection}{\Roman{section}.}

\renewcommand{\theequation}{\thesection\arabic{equation}}
\@addtoreset{equation}{section}
\makeatother

\font\msym=msbm10

\def\Z{{\mathop{\hbox{\msym\char '132}}}}

\def\tx{\tilde{{\rm x}}}

\def\da{\dot{\alpha}} 
\def\db{\dot{\beta}} 
 
\def\bt{\bar{\theta}}

\def\B{{\cal B}}

\def\E{{\cal E}}

\begin{document}
\begin{titlepage}
\title{\vskip -60pt
\vskip 30pt
\textbf{A Study of  Two-/One-form Superfields}
\\~~\\~~~\\ ~~~\\~~~ \\}
\author{Jeong-Hyuck Park\thanks{E-mail address:\,\,jhp@kias.re.kr}}
\date{}
\maketitle
\vspace{-1.0cm}
\begin{center}
\textit{School of Physics, Korea Institute for Advanced Study}\\
\textit{207-43 Cheongryangri-dong, Dongdaemun-gu}\\
\textit{Seoul 130-012, Korea}\\
~~~\\
~~~\\
~~~\\
~~~\\
~~~\\
\end{center}
\begin{abstract}
\noindent   
We  study  two supersymmetric toy models  of  a $k$-form superfield, $k=2,1$ 
separately.   
By ``solving'' Jacobi   identities, we show 
that each model is completely solvable at off-shell level, 
possesses a severely constrained kinematics,  
and gives a   rigid representation 
of the supersymmetry algebra.   
This study of the toy models is motivated by our  
reanalysis on  supersymmetry algebras  in  gauge theories where we discuss   
internal symmetry generators  carrying  spacetime/spinor indices.
\end{abstract}
~\\
{\small
\begin{flushleft} 
~~~~~~~~~\textit{PACS}: 11.30.Pb, 11.15.-q, 12.60.Jv \\
\end{flushleft}}

\thispagestyle{empty}
\end{titlepage}
\newpage
\baselineskip24pt
\section{Introduction}
Coleman-Mandula theorem~\cite{PR1591251}  has been a cornerstone to construct possible supersymmetry algebras in various dimensions. The theorem states that any group of bosonic symmetries of the \textbf{s}-matrix in relativistic field theory is the direct product of of the Poincar\'{e} group with an internal 
symmetry group. Consequently  
supersymmetry algebra was conceived as a $\Z_{2}$-graded algebra 
of which the bosonic part needed to satisfy the theorem~\cite{NPB88257,sohnius,strathdee}. In particular the internal symmetry generators do not carry any spacetime index. One of the assumptions necessary to prove the theorem 
was that the theory describes massive point-like particles. 
For massless theories it is well known that  
the Poincar\'{e} group can be extended to the conformal group, while in the presence of a $p$-dimensional extended object, or $p$-brane, it was pointed out by Azc\'{a}rraga \textit{et al.} that 
the supersymmetry algebra can admit  a $p$-form central 
charge~\cite{PRL632443,9507048}. 
This discovery initiated some studies of the possible central extensions of supersymmetry algebra by adding charges with 
spacetime indices to the ordinary supersymmetry algebra~\cite{9604139,9711116,9903241}. \newline

It has been well known  that in gauge theories the commutator  of  two 
supersymmetry transformations  contains a  gauge transformation as well as a translation~\cite{peterwest,NPB264653,9711161}.   For example, regarding four-dimensional 
${\cal N}=1$  supersymmetric Maxwell theory,  
there exist essentially two known superfield formulations: One is to introduce a real 
scalar superfield or vector superfield and impose the Wess-Zumino gauge condition.  
The other is to start with a fermionic chiral superfield, $W_{\alpha}$,  
and impose a certain reality 
constraint, $D^{\alpha}W_{\alpha}=\bar{D}_{\da}\bar{W}^{\da}$. 
In the former approach, the gauge field appears explicitly in the superfield expansion 
and the  supersymmetry algebra closes on a   gauge 
transformation and a translation, though the original supersymmetry algebra in the superfield formalism closes only on a translation. This is the price for adopting  the 
Wess-Zumino gauge.  
On the other hand in the latter approach,  only  the 
field strength appears in the superfield expansion and  the existence of the gauge field follows  implicitly as the exterior derivative of the field strength vanishes, so that the supersymmetry algebra closes on a translation alone\footnote{ The latter approach is demonstrated  in section~\ref{REVIEW} for the analysis of  $6~\!\mbox{D}$ tensor multiplet theory.}.\newline

The main contents of the present paper are twofold: We first revisit  
supersymmetry algebras  in two   gauge theories,   
$6~\!\mbox{D}$   $(1,0)$ tensor multiplet   containing $B_{\mu\nu}$  and   
$4~\!\mbox{D}$  ${\cal N}=1$  supersymmetric Maxwell theory  
containing $A_{\mu}$.  Analysing  the commutators of  
super charges   in each theory,  
we conceive the notion of \textit{gauge charges} which   carry  spacetime/spinor 
indices and generate local gauge transformations.   
Since both theories are massless and describe free point-like particles,   neither 
Coleman-Mandula theorem is  applicable nor solitonic extended objects are  
present.   
Hence this is  another type of  origin for internal symmetry charges which carry 
spacetime/spinor indices. \newline

When there are infinitely many generators in a supersymmetry algebra, the  
superfield formalism is  not practical anymore, since the expansion of any superfield 
does not terminate at a finite order resulting in infinitely many component fields. 
Motivated by the observation above, in the rest of the paper, we analyse  two off-shell  
supersymmetric toy models of  a $k$-form superfield in $2k+2$ dimensions, $k=2,1$,  
separately.  We let the relevant supersymmetry algebra consist of supercharges and translations only.     By ``solving'' Jacobi   identities, we show 
that each model is completely solvable at off-shell level, 
possesses a severely constrained kinematics  
in the sense that all the component fields are at most quadratic in spacetime coordinates, $x$, and
that  it is possible to determine how the coefficients  in the expansion of each component field in $x$ transform to another under supersymmetry transformations, i.e. these coefficients form a ``rigid'' representation of the supersymmetry algebra\footnote{Cf. Ref. 
\cite{ZPC18229}, where ``soft gauge algebra''  is defined to have field-dependent and hence spacetime-dependent structure ``constants''.}.   \newline

The organization of the present paper is as follows.  
Section \ref{6DTENSOR} contains the reanalysis on $6~\!\mbox{D}$  $(1,0)$ tensor multiplet. We discuss the gauge charges there.   In section \ref{TENSOR2}, we study  the $6~\!\mbox{D}$  $(N,0)$  toy model of a two-form superfield and solve the  model  completely.  Section \ref{YM} deals with a parallel analysis on $4~\!\mbox{D}$ ${\cal N}=1$ supersymmetric Maxwell theory and  a   toy model of a one-form superfield.

\section{ $6~\!\mbox{D}$  $(1,0)$ Tensor Multiplet\label{6DTENSOR} }
\subsection{Off-shell Superfield Formalism\label{REVIEW}}
Here  we reconstruct the off-shell $6~\!\mbox{D}$ $(1,0)$ tensor multiplet within a 
superfield formalism in an algebraic way.  
We also exhibit  an explicit expression for the corresponding 
superfield.\newline

The standard six-dimensional $(N,0)$ supersymmetry algebra is given by 
\begin{equation}
\{Q^{i}_{\alpha},Q^{j}_{\beta}\}=2\E^{ij}\gamma^{A}_{\alpha\beta}P_{A}\,,
\label{susy}
\end{equation}
where  $\E^{ij}$ is a $2N\times 2N$ 
anti-symmetric matrix governing the symplectic structure
\begin{equation}
\E^{ij}=\left(\begin{array}{cc}
0&1\\
-1&0
\end{array}\right)\,,
\end{equation}
with the inverse, 
$\bar{\E}_{ij}$, $\E^{ij}\bar{\E}_{jk}=\delta^{i}{}_{k}$, and super charges, 
$Q^{i}_{\alpha},1\leq i\leq 2N,1\leq \alpha\leq 4$, satisfy the   
pseudo-Majorana  condition  
\begin{equation}
\bar{Q}_{i}=Q^{i\dagger}\gamma^{0}=(Q^{j})^{t}\bar{\E}_{ji}\,.
\label{pseudoM1}
\end{equation}
The corresponding superspace has coordinates, 
$z^{M}=(x^{A},\theta^{i\alpha})$, where Grassmann variable, $\theta^{i}$, satisfies $\bar{\theta}_{i}=\theta^{i\dagger}\tilde{\gamma}^{0}
=(\theta^{j})^{t}\bar{\E}_{ji}$ so that $\bar{\theta}_{i}Q^{i}=\bar{Q}_{i}\theta^{i}$ is real.\newline

Following Howe \textit{et al.} \cite{NPB221331}, to obtain $(1,0)$ tensor multiplet,  
we consider a real scalar superfield, $\Phi(z)$, 
subject to $D^{(i}_{\alpha}D^{j)}_{\beta}\Phi(z)=0$ or equivalently\footnote{$(\,),\,[\,]$ mean  symmetrizing, anti-symmetrizing 
indices with ``strength one''.}\footnote{We note that 
$D^{(i}_{\alpha}D^{j)}_{\beta}=D^{(i}_{[\alpha}D^{j)}_{\beta]}$.}

\begin{equation}
\{Q^{(i}_{\alpha},[Q^{j)}_{\beta},\phi]\}=0\,,
\label{ijsym}
\end{equation}
where $\phi(x)=\Phi(x,0)$.\newline

If we define 
\begin{equation}
\psi^{i}_{\alpha}\equiv i[Q^{i}_{\alpha},\phi]\,,
\label{QQQphi}
\end{equation}
then $\psi^{i}_{\alpha}$ satisfies the pseudo-Majorana condition 
along with $Q^{i}_{\alpha}$
\begin{equation}
\bar{\psi}_{i}=\psi^{i\dagger}\gamma^{0}=(\psi^{j})^{t}\bar{\E}_{ji}\,.
\label{pseudoM2}
\end{equation} 
From eq.(\ref{ijsym}) and using the fact that 
$\gamma^{A},\gamma^{[B}\tilde{\gamma}^{C}\gamma^{D]}$ form a basis of 
$4\times 4$ matrices, one can derive an expression for $\{Q^{i}_{\alpha},\psi^{j}_{\beta}\}$. After imposing a Jacobi identity 
\begin{equation}
\{Q^{i}_{\alpha},[Q^{j}_{\beta},\phi]\}+\{Q^{j}_{\beta},[Q^{i}_{\alpha},\phi]\}
=[\{Q^{i}_{\alpha},Q^{j}_{\beta}\},\phi]=2\E^{ij}\gamma^{A}_{\alpha\beta}[P_{A},\phi]\,,
\end{equation}
we get\footnote{$P_{A}$ acts on fields as a partial derivative in a standard way, $[P_{A},\phi]=-i\partial_{A}\phi$.}
\begin{equation}
\{Q^{i}_{\alpha},\psi^{j}_{\beta}\}=
\E^{ij}(\gamma^{A}\partial_{A}\phi+\textstyle{\frac{1}{4}}
\gamma^{[A}\tilde{\gamma}^{B}\gamma^{C]}H_{ABC})_{\alpha\beta}\,,
\label{QQQpsi}
\end{equation}
where, due to eq.(\ref{Due}), $H_{ABC}$ is self-dual  
\begin{equation}
H_{ABC}=\textstyle{\frac{1}{3!}}\epsilon_{ABC}{}^{DEF}H_{DEF}\,,
\label{self}
\end{equation}
and satisfies from eq.(\ref{tr6})
\begin{equation}
\E^{ij}H_{ABC}=-
\textstyle{\frac{1}{12}}(\tilde{\gamma}_{[A}\gamma_{B}\tilde{\gamma}_{C]})^{\alpha\beta}\{Q^{i}_{\alpha},\psi^{j}_{\beta}\}\,.
\end{equation}
Further, with eq.(\ref{hermiticity})  
pseudo-Majorana conditions~(\ref{pseudoM1},\ref{pseudoM2}) imply  
that $H_{ABC}$ is real.\newline

Using Jacobi identity one can write
\begin{equation}
\begin{array}{ll}
-12\E^{ij}[Q^{k}_{\alpha},H_{ABC}]&=(\tilde{\gamma}_{[A}\gamma_{B}\tilde{\gamma}_{C]})^{\gamma\delta}[Q^{k}_{\alpha},\{Q^{i}_{\gamma},\psi^{j}_{\delta}\}]\\
{}&{}\\
&=(\tilde{\gamma}_{[A}\gamma_{B}\tilde{\gamma}_{C]})^{\gamma\delta}\left(
2\E^{ki}\gamma^{D}_{\alpha\gamma}[P_{D},\psi^{j}_{\delta}]
-[Q^{i}_{\gamma},\{Q^{k}_{\alpha},\psi^{j}_{\delta}\}]\right)\\
{}&{}\\
{}&=\E^{kj}\left(i(\gamma^{D}\tilde{\gamma}_{[A}\gamma_{B}\tilde{\gamma}_{C]}
\partial_{D}\psi^{i})_{\alpha}
-\textstyle{\frac{1}{4}}(\gamma^{[D}\tilde{\gamma}^{E}\gamma^{F]}
\tilde{\gamma}_{[A}\gamma_{B}\tilde{\gamma}_{C]})_{\alpha}{}^{\beta}
[Q^{i}_{\beta},H_{DEF}]\right)\\
{}&{}\\
{}&~~~
-2i\E^{ki}(\gamma^{D}\tilde{\gamma}_{[A}\gamma_{B}\tilde{\gamma}_{C]}
\partial_{D}\psi^{j})_{\alpha}\,.
\end{array}
\label{QQQHpre}
\end{equation}
Choosing $j=k$ gives
\begin{equation}
[Q^{k}_{\alpha},H_{ABC}]=-i\textstyle{\frac{1}{6}}(
\gamma^{D}\tilde{\gamma}_{[A}\gamma_{B}\tilde{\gamma}_{C]}
\partial_{D}\psi^{k})_{\alpha}\,,
\label{QQQH}
\end{equation}
which is in fact  equivalent to eq.(\ref{QQQHpre}) itself.\newline

With eqs.(\ref{QQQpsi},\ref{QQQH},\ref{LEQ},\ref{epep6}) Jacobi identity gives 
\begin{equation}
\begin{array}{ll}
\partial_{D}H_{ABC}&=i\textstyle{\frac{1}{16}}\bar{\E}_{ij}
\tilde{\gamma}^{\alpha\beta}_{D}\left(
\{Q^{i}_{\alpha},[Q^{j}_{\beta},H_{ABC}]\}+
\{Q^{j}_{\beta},[Q^{i}_{\alpha},H_{ABC}]\}\right)\\
{}&{}\\
{}&=2\partial_{[A}H_{BC]D}-\eta_{D[A}\partial^{E}H_{BC]E}+
\textstyle{\frac{1}{6}}\epsilon_{ABC}{}^{EFG}\partial_{E}H_{FGD}\,.
\end{array}
\label{JdH}
\end{equation}
With the self-duality of $H_{ABC}$, using  eq.(\ref{epep6}), this becomes
\begin{equation}
\partial_{D}H_{ABC}=3\partial_{[A}H_{BC]D}-3\eta_{D[A}\partial^{E}H_{BC]E}\,,
\end{equation}
and hence
\begin{eqnarray}
\partial^{C}H_{ABC}=0\,,\label{EMH}\\
{}\nonumber\\
\partial_{[A}H_{BCD]}=0\,.\label{poincare}
\end{eqnarray}
The latter implies that $H_{ABC}$ is a field strength of a certain two-form tensor at least locally
\begin{equation}
H_{ABC}=\partial_{[A}B_{BC]}\,.
\label{HBre}
\end{equation}
Hence the existence of a two-form gauge field follows only implicitly as 
the exterior derivative of the three-form tensor vanishes, and 
$\phi,\psi^{i},H_{ABC}$ form a representation of the standard six-dimensional $(1,0)$ supersymmetry algebra~(\ref{susy}) consisting of $Q,P$ only.  It is worth to   note that  eq.(\ref{EMH}) follows from eq.(\ref{HBre}) and 
the self-duality of $H_{ABC}$.\newline

Constraints on $\phi,\,\psi^{i}$ can be obtained in a similar fashion. From eqs.(\ref{QQQphi},\ref{QQQpsi},\ref{QQQH}) we get
\begin{equation}
[Q^{k}_{\gamma},\{Q^{i}_{\alpha},\psi^{j}_{\beta}\}]=-i\E^{ij}(\gamma^{A}_{\alpha\beta}\partial_{A}\psi^{k}_{\gamma}+\gamma^{A}_{\beta\gamma}\partial_{A}\psi^{k}_{\alpha}-\gamma^{A}_{\gamma\alpha}\partial_{A}\psi^{k}_{\beta})\,,
\end{equation}
so that with Jacobi identity
\begin{equation}
[\{Q^{i}_{\alpha},Q^{k}_{\gamma}\},\psi^{j}_{\beta}]=-i\E^{ik}(\gamma^{A}_{\beta\gamma}\partial_{A}\psi^{j}_{\alpha}+\gamma^{A}_{\alpha\beta}\partial_{A}\psi^{j}_{\gamma}-\gamma^{A}_{\gamma\alpha}\partial_{A}\psi^{j}_{\beta})\,.
\end{equation}
From the supersymmetry algebra this must be equal to $-2i\E^{ik}\gamma^{A}_{\alpha\gamma}\partial_{A}\psi^{j}_{\beta}$, and hence
\begin{equation}
\gamma^{A}_{[\alpha\beta}\partial_{A}\psi^{j}_{\gamma]}=0\,.
\end{equation}
Equivalently from eq.(\ref{gtg})
\begin{equation}
\tilde{\gamma}^{A}\partial_{A}\psi^{j}=0\,.
\label{CPSI}
\end{equation}
Furthermore this implies  
$\tilde{\gamma}^{A}{}^{\alpha\beta}\partial_{A}
\{Q^{i}_{\alpha},\psi^{j}_{\beta}\}=0$, and hence with eq.(\ref{QQQpsi})
\begin{equation}
\Box\,\phi=0\,.
\end{equation}

Now we can write  superfield
\begin{equation}
\begin{array}{ll}
\Phi(z)&=\displaystyle{e^{i\bt_{i}Q^{i}}\phi(x)e^{-i\bt_{i}Q^{i}}}\\
{}&{}\\
{}&=\phi+\bar{\theta}_{i}\psi^{i}+i\textstyle{\frac{1}{8}}
\bar{\theta}_{i}\gamma^{[A}\tilde{\gamma}^{B}\gamma^{C]}\theta^{i}H_{ABC}
+i\textstyle{\frac{1}{3}}\bar{\theta}_{i}\gamma^{A}\theta^{j}\bar{\theta}_{j}\partial_{A}\psi^{i}+\textstyle{\frac{1}{12}}\bar{\theta}_{i}\gamma^{A}\theta^{j}
\bar{\theta}_{j}\gamma^{B}\theta^{i}\partial_{A}\partial_{B}\phi\\
{}&{}\\
{}&~~~+\mbox{higher~order~terms~containing~derivatives}\,,
\end{array}
\end{equation}
where  the higher order terms which terminate at eighth order  
can be obtained from \newline 
eqs.(\ref{QQQphi},\ref{QQQpsi},\ref{QQQH}).\newline

\subsection{Charges with Spacetime Indices : On-shell\label{TENSOR1}}
The   supersymmetry transformation rule for the two-form tensor field 
was first  written by Bergshoeff \textit{et  al.} at   on-shell level~\cite{NPB264653,turin} as
\begin{equation}
\delta B_{AB}=\bar{\varepsilon}_{i}\gamma_{[A}\tilde{\gamma}_{B]}\psi^{i}\,.
\label{deltaB}
\end{equation} 
From their results, by identifying
\begin{equation}
\delta\,\mbox{Field}=i[\bar{\varepsilon}_{i}Q^{i},\mbox{Field}]\,,
\end{equation}
one may introduce supersymmetry charges, 
$Q^{i}_{\alpha},1\leq i\leq 2,$  
such that
\begin{subeqnarray}
\label{compsusy}
[Q^{i}_{\alpha},B_{AB}]&=&
-i(\gamma_{[A}\tilde{\gamma}_{B]}\psi^{i})_{\alpha}\,,\label{QBpre}\\
{}\nonumber\\
\{Q^{i}_{\alpha},\psi^{j}_{\beta}\}&=&\E^{ij}(\gamma^{A}\partial_{A}\phi+\textstyle{\frac{1}{4}}
\gamma^{[A}\tilde{\gamma}^{B}\gamma^{C]}H_{ABC})_{\alpha\beta}\,,\\
{}\nonumber\\
{}[Q^{i}_{\alpha},\phi]&=&-i\psi^{i}_{\alpha}\,.
\end{subeqnarray}
This is  compatible with eqs.(\ref{QQQphi},\ref{QQQpsi},\ref{QQQH}).   
We also note that eq.(\ref{compsusy}) is  consistent with 
the pseudo-Majorana conditions~(\ref{pseudoM1},\ref{pseudoM2}), which can be shown using eq.(\ref{hermiticity}).\newline

From eq.(\ref{compsusy}) with Jacobi identities
\begin{subeqnarray}
&[\{Q^{i}_{\alpha},Q^{j}_{\beta}\},\mbox{Boson}]=
\{Q^{i}_{\alpha},[Q^{j}_{\beta},\mbox{Boson}]\}+
\{Q^{j}_{\beta},[Q^{i}_{\alpha},\mbox{Boson}]\}\,,~~&\label{JB}\\
{}\nonumber\\
&[\{Q^{i}_{\alpha},Q^{j}_{\beta}\},\mbox{Fermion}]=
[Q^{i}_{\alpha},\{Q^{j}_{\beta},\mbox{Fermion}\}]+
[Q^{j}_{\beta},\{Q^{i}_{\alpha},\mbox{Fermion}\}]\,,\label{JF}&
\end{subeqnarray}
one can calculate the anti-commutator of super charges. 
Using eq.(\ref{LEQ}) we get
\begin{subeqnarray}
\label{offshell}
&\begin{array}{ll}
[\{Q^{i}_{\alpha},Q^{j}_{\beta}\},B_{AB}]&=-2i
\E^{ij}\partial_{C}B_{AB}\gamma^{C}_{\alpha\beta}
+3i\E^{ij}H_{-ABC}\gamma^{C}_{\alpha\beta}\\
{}&{}\\
{}&~~~-2i\E^{ij}
(\partial_{A}B_{BC}-\partial_{B}B_{AC}+\eta_{CA}\partial_{B}\phi-
\eta_{CB}\partial_{A}\phi)\gamma^{C}_{\alpha\beta}\,,\end{array}\\
&\nonumber\\
&\nonumber\\
&\begin{array}{ll}
[\{Q^{i}_{\alpha},Q^{j}_{\beta}\},\psi^{k}_{\gamma}]&=
i\E^{jk}(\gamma^{A}_{\alpha\beta}\partial_{A}\psi^{i}_{\gamma}-\gamma^{A}_{\beta\gamma}\partial_{A}\psi^{i}_{\alpha}-\gamma^{A}_{\gamma\alpha}\partial_{A}\psi^{i}_{\beta})-(i\leftrightarrow j)\\
{}&{}\\
{}&=-2i\E^{ij}\partial_{A}\psi^{k}_{\gamma}\gamma^{A}_{\alpha\beta}+i\textstyle{\frac{1}{2}}\E^{ij}(\gamma_{A}\tilde{\gamma}^{B}\partial_{B}\psi^{k})_{\gamma}\gamma^{A}_{\alpha\beta}\,,\end{array}~~\label{N=1}\\
&\nonumber\\
&\nonumber\\
&\begin{array}{ll}
[\{Q^{i}_{\alpha},Q^{j}_{\beta}\},\phi]&=
-2i\E^{ij}\partial_{A}\phi\gamma^{A}_{\alpha\beta}\,,~~~~~~~~~~~~~~~~~~~~~~~~~~~~~~~~~~~~~~~~~~~~
\end{array}
\end{subeqnarray}
where  
\begin{equation}
H_{\pm ABC}\equiv\textstyle{\frac{1}{2}}(H_{ABC}\pm\textstyle{\frac{1}{3!}}\epsilon_{ABC}{}^{DEF}H_{DEF})\,.
\end{equation}
The equality on the second line of eq.(\ref{N=1}) holds since $N=1$, i.e. 
$1\leq i,j\leq 2$.\newline

Now we introduce a vector valued Hermitian charge, $Z_{A}$, such that
\begin{equation}
\begin{array}{cc}
\multicolumn{2}{c}{
[Z_{C},B_{AB}]=-i(\partial_{A}B_{BC}-\partial_{B}B_{AC}+\eta_{CA}\partial_{B}\phi-\eta_{CB}\partial_{A}\phi)\,,}\\
{}&{}\\
{}[Z_{C},\psi^{i}_{\alpha}]=0\,,~~~~&~~~~[Z_{C},\phi]=0\,.
\end{array}
\end{equation}
Direct calculation gives
\begin{equation}
[Z_{A},H_{BCD}]=0\,,
\end{equation}
and hence $Z_{A}$ is a vector valued charge which generates a local gauge 
transformation. We also note that $Z_{A}$ is effectively null when it 
acts on $\psi^{i}_{\alpha},\phi,H_{ABC}$. \newline

\indent With the following constraints or ``equations of motion''\footnote{Contrary to the known, we do not need any constraint on the scalar field, $\phi$.} 
\begin{subeqnarray}
&H_{-ABC}=0~~\mbox{:\,self-duality}\,,&\label{self2}\\
{}\nonumber\\
&\tilde{\gamma}^{A}\partial_{A}\psi^{i}=0\,,&\label{EMpsi}
\end{subeqnarray}
eq.(\ref{offshell}) realizes the following novel modified $(1,0)$ supersymmetry algebra in six-dimensions
\begin{equation}
\{Q^{i}_{\alpha},Q^{j}_{\beta}\}=2\E^{ij}
\gamma^{A}_{\alpha\beta}(P_{A}+Z_{A})\,.
\label{anticomQQ}
\end{equation}
We note from eqs.(\ref{QBpre},\ref{6gamma})
\begin{equation}
\begin{array}{ll}
[Q^{i}_{\alpha},H_{ABC}]&=-i(\gamma_{[A}\tilde{\gamma}_{B}\partial_{C]}\psi^{i})_{\alpha}\\
{}&{}\\
{}&=-i\textstyle{\frac{1}{6}}(\gamma^{D}\tilde{\gamma}_{[A}\gamma_{B}\tilde{\gamma}_{C]}\partial_{D}\psi^{i}+\gamma_{[A}\tilde{\gamma}_{B}\gamma_{C]}\tilde{\gamma}^{D}\partial_{D}\psi^{i})_{\alpha}\,.
\end{array}
\end{equation}
This shows from eq.(\ref{Due}) that the self-duality condition~(\ref{self}) is also satisfied for the expression of $[Q^{i}_{\alpha},H_{ABC}]$ with  
the equation of motion~(\ref{EMpsi})
\begin{equation}
[Q^{i}_{\alpha},H_{-ABC}]=-i\textstyle{\frac{1}{6}}
(\gamma_{[A}\tilde{\gamma}_{B}\gamma_{C]}\tilde{\gamma}^{D}\partial_{D}\psi^{i})_{\alpha}=0\,.
\end{equation}

Similar analysis based on Jacobi identities can give other commutators/anti-commutators of $Q^{i}_{\alpha},P_{A},Z_{A}$.  First we find
\begin{equation}
\begin{array}{ccc}
[P_{A},Q^{i}_{\alpha}]=0\,,~~~~&~~~~[P_{A},P_{B}]=0\,,
~~~~&~~~~[P_{A},Z_{B}]=0\,.
\end{array}
\label{Pcommutes}
\end{equation}
If we denote gauge charges ${\cal T}_{(n)}$ which are 
 $n$ times multi-commutators of $Q^{i}_{\alpha},Z_{A}$ containing at least one $Z_{A}$, i.e.
\begin{equation}
\begin{array}{l}
{\cal T}_{A_{1}A_{2}\cdots A_{n}}=[Z_{A_{1}},[Z_{A_{2}},\cdots,
[Z_{A_{n-1}},Z_{A_{n}}]\cdots ]\equiv Z_{A_{1}A_{2}\cdots A_{n}}\,,\\
{}\\
{\cal T}^{i}_{\alpha}{}^{j}_{\beta}{}_{A}=\{Q^{i}_{\alpha},[Q^{j}_{\beta},Z_{A}]\}\,,~~~~~~~~~~~~~~~\mbox{etc.}
\end{array}
\end{equation}
then one can show, by induction on $n$, that gauge charges, 
${\cal T}_{(n)}$,  act only on the two-form tensor, $B_{AB}$,
\begin{equation}
\begin{array}{ccc}
[{\cal T}_{(n)},\psi^{i}_{\alpha}\}=0\,,~~~&~~~
[{\cal T}_{(n)},\phi]=0\,,~~~&~~~
[{\cal T}_{(n)},H_{ABC}]=0\,,
\end{array}
\end{equation}
and ${\cal T}_{(n)}$  commutes with $P_{A}$
\begin{equation}
[{\cal T}_{(n)},P_{A}]=0\,.
\label{Pcommutesagain}
\end{equation}
In particular, direct calculation gives
\begin{subeqnarray}
&[[Z_{C},Q^{i}_{\alpha}],B_{AB}]=(\gamma_{C}(\tilde{\gamma}_{A}\partial_{B}-
\tilde{\gamma}_{B}\partial_{A})\psi^{i})_{\alpha}\,,\\
{}\nonumber\\
&{}[Z_{CD},B_{AB}]=\partial_{A}(\partial_{D}B_{BC}-\partial_{C}B_{BD}+\eta_{DB}\partial_{C}\phi-\eta_{CB}\partial_{D}\phi)-(A\leftrightarrow B)\,,\\
{}\nonumber\\
&{}[Z_{C_{1}C_{2}\cdots C_{n}},B_{AB}]=\displaystyle{\prod^{n-2}_{j=1}}
(-i\partial_{C_{j}})\left(\partial_{A}(\partial_{C_{n}}B_{BC_{n-1}}-
\partial_{C_{n-1}}B_{BC_{n}})-(A\leftrightarrow B)\right)\,.
\end{subeqnarray}
Since  $[{\cal T}_{(n)},Q_{\alpha}^{i}\}\neq 0$, 
gauge charges are not central.\newline

We note that there exist infinitely many gauge charges,  ${\cal T}_{(n)}$, and this has to do with the fact    that 
 there exist infinite degrees of freedom for  local gauge 
transformations.   It is worth to  note that 
Jacobi identities may reduce the number of independent charges such as
\begin{equation}
{\cal T}^{i}_{\alpha}{}^{j}_{\beta}{}_{A}+
{\cal T}^{j}_{\beta}{}^{i}_{\alpha}{}_{A}
=2\E^{ij}\gamma^{B}_{\alpha\beta}Z_{BA}\,.
\end{equation}

\section{$6~\!\mbox{D}$ Toy Model of Two-form Superfield : Off-shell\label{TENSOR2}}
In this section we investigate the consequences of setting  $Z_{A}=0$, by considering a $6~\!\mbox{D}$ off-shell toy model of a two-form superfield. 
We  assume the supersymmetry 
transformation of the lowest component field of the superfield  is identical to that of the two-form gauge field  
in $6~\!\mbox{D}$ tensor multiplet~(\ref{QBpre}), i.e.
\begin{equation}
[Q^{i}_{\alpha},B_{AB}]=-i(\gamma_{[A}\tilde{\gamma}_{B]}\psi^{i})_{\alpha}\,,
\label{QB}
\end{equation}
and require our toy model to be subject to the standard $(N,0)$ 
supersymmetry algebra consisting of 
$Q^{i}_{\alpha},P_{A}$ alone, i.e.  eq.(\ref{susy})
\begin{equation}
\{Q^{i}_{\alpha},Q^{j}_{\beta}\}=2\E^{ij}\gamma^{A}_{\alpha\beta}P_{A}\,,
\label{susy2}
\end{equation}
where $1\leq i,j\leq 2N$ and $N$ is an arbitrary natural number. We note that eq.(\ref{QB}) is of the most general form without any derivative and gravitinos.  \newline

Provided with  these two equations~(\ref{QB},\ref{susy2}) only, without any ``equation of motion'',  
using some properties of gamma matrices in six-dimensions   
we are going to show that the model is completely solvable at off-shell level, 
possesses a severely constrained kinematics  as 
\begin{itemize}
\item $B_{AB},\psi^{i}_{\alpha}$ alone form a super multiplet 
without introducing any scalar field.
\item The field strength of the two-form field is automatically self-dual.
\item $B_{AB}$ is at most linear in spacetime coordinates, $x$, and it has the following  explicit form 
\begin{equation}
B_{AB}=\B_{A}x_{B}-\B_{B}x_{A}+\B_{ABC}x^{C}+\B_{AB}\,,
\label{27}
\end{equation}
where
\begin{equation}
\begin{array}{cc}
\B_{AB}=\B_{[AB]}\,,~~~~~&~~~~
\B_{ABC}=\textstyle{\frac{1}{3!}}\epsilon_{ABC}{}^{DEF}\B_{DEF}\,.
\end{array}
\label{calB}
\end{equation}
\item The spinor field, $\psi^{i}_{\alpha}$, 
is a spacetime independent constant field, i.e. there remain only zero modes and with $\B_{\alpha}{}^{\beta}\equiv 
\B_{AB}(\gamma^{[A}\tilde{\gamma}^{B]})_{\alpha}{}^{\beta}$ it can be written as
\begin{equation}
\psi^{i}_{\alpha}=
-i\textstyle{\frac{1}{30}}[Q^{i}_{\beta},\B_{\alpha}{}^{\beta}]\,.
\label{psiQB}
\end{equation}
\item The supersymmetry transformation rule for $\psi^{i}_{\alpha}$ is
\begin{equation}
\{Q^{i}_{\alpha},\psi^{j}_{\beta}\}=\E^{ij}
(\textstyle{\frac{1}{5}}\partial_{A}B^{A}{}_{B}\gamma^{B}+
\textstyle{\frac{1}{12}}
H_{ABC}\gamma^{[A}\tilde{\gamma}^{B}\gamma^{C]})_{\alpha\beta}\,.
\label{8}
\end{equation}
\item The supersymmetry algebra~(\ref{susy2}) is represented by
\begin{subeqnarray}
\label{QBs}
&[Q^{i}_{\alpha},\B_{A}]=0\,,~~~~~~~~~~[Q^{i}_{\alpha},\B_{ABC}]=0\,,&\label{QB0}\\
&{}&\nonumber\\
&[Q^{i}_{\alpha},\B_{AB}]=-i(\gamma_{[A}\tilde{\gamma}_{B]}\psi^{i})_{\alpha}\,,&
\label{QBpsi}\\
&{}&\nonumber\\
&\{Q^{i}_{\alpha},\psi^{j}_{\beta}\}=\E^{ij}(-\B^{A}\gamma_{A}
+\textstyle{\frac{1}{12}}\B_{ABC}\gamma^{[A}\tilde{\gamma}^{B}\gamma^{C]}
)_{\alpha\beta}\,,&\label{Qpsi}
\end{subeqnarray}
\begin{subeqnarray}
\label{PBs}
&[P_{C},\B_{A}]=0\,,~~~~~~~~~~[P_{C},\B_{ABC}]=0\,,&\label{PB0}\\
&{}&\nonumber\\
&[P_{C},\B_{AB}]=-i(\B_{A}\eta_{BC}-\B_{B}\eta_{CA}+\B_{ABC})\,,&\label{PBB}\\
&{}&\nonumber\\
&[P_{C},\psi^{i}]=0\,.\label{Ppsi}
\end{subeqnarray}
\end{itemize}

{}~\newline
\indent\textit{Proof}\newline
\newline
With $B_{\alpha}{}^{\beta}\equiv 
B_{AB}(\gamma^{[A}\tilde{\gamma}^{B]})_{\alpha}{}^{\beta}$, eq.(\ref{QB}) is equivalent from eq.(\ref{complete}) to 
\begin{equation}
[Q^{i}_{\alpha},B_{\beta}{}^{\gamma}]=8i\delta_{\alpha}{}^{\gamma}\psi^{i}_{\beta}-2i\delta_{\beta}{}^{\gamma}\psi^{i}_{\alpha}\,.
\label{QB2}
\end{equation}
Jacobi identity which is crucial in our calculation is with eq.(\ref{susy2})
\begin{equation}
\{Q^{i}_{\alpha},[Q^{j}_{\beta},B_{\gamma}{}^{\delta}]\}+
\{Q^{j}_{\beta},[Q^{i}_{\alpha},B_{\gamma}{}^{\delta}]\}=
2\E^{ij}\gamma^{A}_{\alpha\beta}[P_{A},B_{\gamma}{}^{\delta}]\,,
\label{preJacobi}
\end{equation}
and hence from eq.(\ref{QB2})
\begin{equation}
\delta_{\gamma}{}^{\delta}\{Q^{i}_{\alpha},\psi^{j}_{\beta}\}+
\delta_{\gamma}{}^{\delta}\{Q^{j}_{\beta},\psi^{i}_{\alpha}\}
-4\delta_{\beta}{}^{\delta}\{Q^{i}_{\alpha},\psi^{j}_{\gamma}\}
-4\delta_{\alpha}{}^{\delta}\{Q^{j}_{\beta},\psi^{i}_{\gamma}\}
=\E^{ij}\gamma^{A}_{\alpha\beta}\partial_{A}B_{\gamma}{}^{\delta}\,.
\label{Jacobi}
\end{equation}
Contracting $\beta,\,\delta$ indices gives
\begin{equation}
15\{Q^{i}_{\alpha},\psi^{j}_{\gamma}\}
+4\{Q^{j}_{\alpha},\psi^{i}_{\gamma}\}
-\{Q^{j}_{\gamma},\psi^{i}_{\alpha}\}
=\E^{ij}\partial_{A}B_{BC}
(\gamma^{[B}\tilde{\gamma}^{C]}\gamma^{A})_{\gamma\alpha}\,.
\label{3}
\end{equation}
Symmetrizing $i,\,j$ indices of this gives
\begin{equation}
19\{Q^{(i}_{\alpha},\psi^{j)}_{\gamma}\}
=\{Q^{(i}_{\gamma},\psi^{j)}_{\alpha}\}\,,
\end{equation}
so that
\begin{equation}
\{Q^{i}_{\alpha},\psi^{j}_{\beta}\}+
\{Q^{j}_{\alpha},\psi^{i}_{\beta}\}=0\,.
\end{equation}
Hence eq.(\ref{3}) becomes
\begin{equation}
11\{Q^{i}_{\alpha},\psi^{j}_{\gamma}\}
+\{Q^{i}_{\gamma},\psi^{j}_{\alpha}\}
=\E^{ij}\partial_{A}B_{BC}
(\gamma^{[B}\tilde{\gamma}^{C]}\gamma^{A})_{\gamma\alpha}\,.
\label{5}
\end{equation}
Using eq.(\ref{plus}) we symmetrize $\alpha,\,\gamma$ indices  to get
\begin{equation}
\{Q^{i}_{\alpha},\psi^{j}_{\gamma}\}+
\{Q^{i}_{\gamma},\psi^{j}_{\alpha}\}
=\textstyle{\frac{1}{6}}\E^{ij}
H_{ABC}(\gamma^{[A}\tilde{\gamma}^{B}\gamma^{C]})_{\alpha\gamma}\,,
\label{6}
\end{equation}
On the other hand, from eq.(\ref{minus}), anti-symmetrizing  
$\alpha,\,\gamma$ indices in eq.(\ref{5}) gives
\begin{equation}
\{Q^{i}_{\alpha},\psi^{j}_{\gamma}\}-
\{Q^{i}_{\gamma},\psi^{j}_{\alpha}\}
=\textstyle{\frac{2}{5}}\E^{ij}\partial_{A}B^{A}{}_{B}
\gamma^{B}_{\alpha\gamma}\,.
\label{7}
\end{equation}
Thus, combining eq.(\ref{6}) and eq.(\ref{7}) gives 
the supersymmetry transformation rule for the spinor field~(\ref{8}). 
Substituting this into the Jacobi identity~(\ref{Jacobi}) gives
\begin{equation}
\textstyle{\frac{8}{5}}\delta_{[\alpha}{}^{\delta}\gamma^{B}_{\beta]\gamma}
\partial_{A}B^{A}{}_{B}
+\textstyle{\frac{2}{3}}\delta_{[\alpha}{}^{\delta}
(\gamma^{[A}\tilde{\gamma}^{B}\gamma^{C]})_{\beta]\gamma}H_{ABC}
+\textstyle{\frac{2}{5}}\delta_{\gamma}{}^{\delta}\gamma^{B}_{\alpha\beta}
\partial_{A}B^{A}{}_{B}
=\gamma^{A}_{\alpha\beta}\partial_{A}B_{\gamma}{}^{\delta}\,.
\label{9}
\end{equation}
Contracting this with $\tilde{\gamma}_{A}^{\alpha\beta}$ gives 
from eq.(\ref{tr2})
\begin{equation}
\partial_{A}B_{\gamma}{}^{\delta}=
(\textstyle{\frac{2}{5}}\partial_{B}B^{BC}
\gamma_{[A}\tilde{\gamma}_{C]}+\textstyle{\frac{1}{6}}H_{BCD}
\gamma^{[B}\tilde{\gamma}^{C}\gamma^{D]}\tilde{\gamma}_{A}
)_{\gamma}{}^{\delta}\,.
\label{10}
\end{equation}
Since this vanishes when we contract $\gamma\,,\delta$ indices,   from      
eqs.(\ref{tr4},\ref{tr6}),  contracting eq.(\ref{10}) with 
$(\gamma^{[B}\tilde{\gamma}^{C]})_{\delta}{}^{\gamma}$   
simplifies eq.(\ref{10}) to the following equivalent formula  
\begin{equation}
\partial_{A}B_{BC}=\textstyle{\frac{1}{5}}(\eta_{AB}\partial_{D}B^{D}{}_{C}-
\eta_{AC}\partial_{D}B^{D}{}_{B})+H_{+ABC}\,.
\label{master}
\end{equation}
\newline

Now we are going to solve eq.(\ref{master}).   
An immediate consequence of eq.(\ref{master}) is that 
$H_{ABC}$ is self-dual~(\ref{self}).
With this self-duality condition, eq.(\ref{master}) becomes
\begin{equation}
10\partial_{A}B_{BC}-5\partial_{B}B_{CA}-5\partial_{C}B_{AB}=
3\eta_{AB}\partial_{D}B^{D}{}_{C}-3\eta_{AC}\partial_{D}B^{D}{}_{B}\,.
\label{13}
\end{equation}
Acting $\partial^{B}$ on this gives
\begin{equation}
7\partial_{A}\partial_{B}B^{B}{}_{C}-5\Box B_{CA}+5
\partial_{C}\partial_{B}B^{B}{}_{A}=0\,.
\label{15}
\end{equation}
Symmetrizing $A,\,C$ indices gives
\begin{equation}
\partial_{A}\partial_{B}B^{B}{}_{C}+\partial_{C}\partial_{B}B^{B}{}_{A}=0\,,
\end{equation}
and hence from eq.(\ref{15}) 
\begin{subeqnarray}
&\partial_{A}\partial_{B}B^{B}{}_{C}=-\textstyle{\frac{5}{2}}\Box B_{AC}\,,&
\label{16}\\
{}\nonumber\\
&\Box\partial_{B}B^{B}{}_{C}=0\,.&\label{16.5}
\end{subeqnarray}
With eq.(\ref{16}), 
acting $\partial_{D}$ on eq.(\ref{13}) and then symmetrizing 
$A,\,B$ indices gives
\begin{equation}
2\partial_{A}\partial_{D}B_{BC}+2\partial_{B}\partial_{D}B_{AC}
=\Box(2\eta_{AB}B_{CD}-\eta_{AC}B_{BD}-\eta_{BC}B_{AD})\,.
\label{18}
\end{equation}
Similarly 
\begin{subeqnarray}
&2\partial_{D}\partial_{A}B_{BC}+2\partial_{B}\partial_{A}B_{DC}
=\Box(2\eta_{DB}B_{CA}-\eta_{DC}B_{BA}-\eta_{BC}B_{DA})\,,&\label{19.a}\\
&{}&\nonumber\\
&2\partial_{D}\partial_{B}B_{AC}+2\partial_{A}\partial_{B}B_{DC}
=\Box(2\eta_{DA}B_{CB}-\eta_{DC}B_{AB}-\eta_{AC}B_{DB})\,.&\label{19.b}
\end{subeqnarray}
Now we add eq.(\ref{19.a}) to eq.(\ref{18}) and then subtract eq.(\ref{19.b}) 
to get
\begin{equation}
\partial_{A}\partial_{D}B_{BC}=\textstyle{\frac{1}{2}}
\Box(\eta_{AB}B_{CD}-\eta_{AC}B_{BD}+\eta_{DB}B_{CA}-\eta_{DC}B_{BA}
+\eta_{AD}B_{BC})\,.
\label{20}
\end{equation}
With eq.(\ref{16.5}) this implies
\begin{equation}
\partial_{A}\Box B_{BC}=\textstyle{\frac{3}{2}}\Box H_{ABC}\,,
\label{21}
\end{equation}
and hence
\begin{equation}
\Box H_{ABC}=\partial_{A}\Box B_{BC}=0\,.
\end{equation}
Thus from eq.(\ref{20})
\begin{equation}
\partial_{A}\partial_{B}\partial_{C}B_{DE}=0\,.
\label{pppB}
\end{equation}
Now if we write
\begin{equation}
\textstyle{\frac{1}{2}}\Box B_{AB}\equiv C_{AB}~~~~\mbox{:~~constant}\,,
\end{equation}
then from eqs.(\ref{20},\ref{pppB}) the quadratic term of $B_{AB}$ is of the form
\begin{equation}
B^{(2)}_{AB}=\textstyle{\frac{1}{2}}C_{AB}x^{2}-C_{AC}x^{C}x_{B}
+C_{BC}x^{C}x_{A}\,,
\end{equation}
so that imposing the self-duality condition~(\ref{self}) gives
\begin{equation}
C_{[AB}x_{C]}=\textstyle{\frac{1}{3!}}\epsilon_{ABC}{}^{DEF}C_{[DE}x_{F]}\,.
\end{equation}
Therefore
\begin{equation}
C_{AB}=0\,,
\end{equation}
and $B_{AB}$ is at most linear in  $x$. From eq.(\ref{master})  
$B_{AB}$ is of the final form~(\ref{27}) 
with coefficients satisfying eq.(\ref{calB}). Due to the explicit form of $B_{AB}$  we get eqs.(\ref{PB0},\ref{PBB}) and  
eq.(\ref{8}) becomes eq.(\ref{Qpsi}).\newline

\indent Now we perform a similar analysis on $\psi^{i}$. A Jacobi identity for this field is
\begin{equation}
[Q^{i}_{\alpha},\{Q^{j}_{\beta},\psi^{k}_{\gamma}\}]+
[Q^{j}_{\beta},\{Q^{i}_{\alpha},\psi^{k}_{\gamma}\}]=
-2i\E^{ij}\gamma^{A}_{\alpha\beta}\partial_{A}\psi^{k}_{\gamma}\,.
\label{28}
\end{equation}
Using eqs.(\ref{L6},\ref{L7}) we get from   eqs.(\ref{QB},\ref{8})
\begin{equation}
[Q^{i}_{\alpha},\{Q^{j}_{\beta},\psi^{k}_{\gamma}\}]=
i\E^{jk}(
\textstyle{\frac{1}{5}}\gamma^{A}_{\beta\gamma}\partial_{A}\psi^{i}_{\alpha}+
\textstyle{\frac{1}{15}}\gamma^{A}_{\gamma\alpha}\partial_{A}\psi^{i}_{\beta}+
\textstyle{\frac{11}{15}}\gamma^{A}_{\alpha\beta}\partial_{A}\psi^{i}_{\gamma}
)\,.
\label{29}
\end{equation}
With this expression, in the case of $i=j$, eq.(\ref{28}) becomes
\begin{equation}
\gamma^{A}_{\beta\gamma}\partial_{A}\psi^{i}_{\alpha}=
-\gamma^{A}_{\alpha\gamma}\partial_{A}\psi^{i}_{\beta}\,,
\end{equation}
and hence $\gamma^{A}_{\alpha\beta}\partial_{A}\psi^{i}_{\gamma}$ is totally 
anti-symmetric for the spinorial indices
\begin{equation}
\Gamma^{i}_{\alpha\beta\gamma}\equiv
\gamma^{A}_{\alpha\beta}\partial_{A}\psi^{i}_{\gamma}=
\Gamma^{i}_{[\alpha\beta\gamma]}\,.
\end{equation}
Now the Jacobi identity~(\ref{28}) becomes
\begin{equation}
\E^{ik}\Gamma^{j}_{\alpha\beta\gamma}-
\E^{jk}\Gamma^{i}_{\alpha\beta\gamma}=
2\E^{ij}\Gamma^{k}_{\alpha\beta\gamma}\,.
\label{31}
\end{equation}
If we consider the case, $i=k$
\begin{equation}
\E^{ij}\Gamma^{i}_{\alpha\beta\gamma}=0\,,
\end{equation}
and hence
\begin{equation}
\partial_{A}\psi^{i}_{\alpha}=0\,,
\end{equation}
$\psi^{i}_{\alpha}$ is a constant field satisfying eq.(\ref{Ppsi}). \newline

\indent Now let's go back to the two-form tensor.  
From eqs.(\ref{QB},\ref{27}) we get
\begin{equation}
[Q^{i}_{\alpha},\B_{A}]x_{B}-[Q^{i}_{\alpha},\B_{B}]x_{A}+
[Q^{i}_{\alpha},\B_{ABC}]x^{C}+[Q^{i}_{\alpha},\B_{AB}]
=-i(\gamma_{[A}\tilde{\gamma}_{B]}\psi^{i})_{\alpha}\,.
\end{equation}
Since $\psi^{i}_{\alpha}$ is  constant  we get eqs.(\ref{QBs},\ref{QBpsi}),  
and from eq.(\ref{QB2}) we get eq.(\ref{psiQB}) as well.\newline

\indent Finally one can check that eqs.(\ref{QBs},\ref{PBs}) are consistent with the supersymmetry algebra~(\ref{susy2}).\newline

\begin{flushright}\textit{Q.E.D.}\end{flushright}

\newpage

\section{$4~\!\mbox{D}$ Super Maxwell Theory and Toy Model\label{YM}}
In this section, similarly to section \ref{6DTENSOR} and \ref{TENSOR2}, we analyse the supersymmetry algebra in $4~\!\mbox{D}$ ${\cal N}=1$ supersymmetric 
Maxwell theory and study a toy model of a one-form superfield.  We obtain similar results.


\subsection{Charges with Spacetime Indices : On-shell\label{YM1}}
As an analogy to eq.(\ref{compsusy}), from  the  known  supersymmetry transformation rules for the  component  fields in  ${\cal N}=1$ on-shell 
supersymmetric Maxwell   multiplet~\cite{buchbinder,peterwest},  we can write\footnote{An auxiliary scalar field was set to be zero  for simplicity.},
\begin{subeqnarray}
{}&[Q_{\alpha},A_{\mu}]=(\sigma_{\mu}\bar{\psi})_{\alpha}\,,~~~~~~~~~~~~~~~~&~~~~
{}[\bar{Q}_{\da},A_{\mu}]=-(\psi\sigma_{\mu})_{\da}\,,\label{QAorig}\\
{}\nonumber\\
{}&\{Q_{\alpha},\psi^{\beta}\}=i\textstyle{\frac{1}{2}}(\sigma^{[\mu}\tilde{\sigma}^{\nu]})_{\alpha}{}^{\beta}F_{\mu\nu}\,,~~~~&~~~~
\{\bar{Q}_{\da},\bar{\psi}^{\db}\}
=i\textstyle{\frac{1}{2}}(\tilde{\sigma}^{[\mu}\sigma^{\nu]})^{\db}{}_{\da}F_{\mu\nu}\,,\\
&{}&{}\nonumber\\
{}&\{Q_{\alpha},\bar{\psi}^{\da}\}=0\,,~~~~~~~~~~~~~~~~~~~~~~~
&~~~~\{\bar{Q}_{\da},\psi^{\alpha}\}=0\,.
\end{subeqnarray}
where  $F_{\mu\nu}=\partial_{\mu}A_{\nu}-\partial_{\nu}A_{\mu}$.\newline
From these, using Jacobi identities and 
eqs.(\ref{fcontract},\ref{fese},\ref{fsig4},\ref{fdde}), direct calculation gives
\begin{subequations}
\begin{equation}
\begin{array}{cc}
\multicolumn{2}{c}{
[\{Q_{\alpha},\bar{Q}_{\da}\},A_{\mu}]=
2iF_{\mu\nu}\sigma^{\nu}_{\alpha\da}\,,}\\
{}&{}\\
{}[\{Q_{\alpha},Q_{\beta}\},A_{\mu}]=0\,,~~~~&~~~~
[\{\bar{Q}_{\da},\bar{Q}_{\db}\},A_{\mu}]=0\,,
\end{array}
\end{equation}
\newline
\begin{equation}
\begin{array}{l}
{}[\{Q_{\alpha},\bar{Q}_{\da}\},\psi^{\beta}]=
-2i\sigma^{\mu}_{\alpha\da}\partial_{\mu}\psi^{\beta}+i(\partial_{\mu}\psi\sigma^{\mu})_{\da}\delta_{\alpha}{}^{\beta}\,,\\
{}\\
{}[\{Q_{\alpha},Q_{\beta}\},\psi^{\gamma}]=i\left(
\delta_{\alpha}{}^{\gamma}(\sigma^{\mu}\partial_{\mu}\bar{\psi})_{\beta}+
\delta_{\beta}{}^{\gamma}(\sigma^{\mu}\partial_{\mu}\bar{\psi})_{\alpha}
\right)\,,\\
{}\\
{}[\{\bar{Q}_{\da},\bar{Q}_{\db}\},\psi^{\alpha}]=0\,,
\end{array}
\end{equation}
\newline
\begin{equation}
\begin{array}{l}
{}[\{Q_{\alpha},\bar{Q}_{\da}\},\bar{\psi}^{\db}]=
-2i\sigma^{\mu}_{\alpha\da}\partial_{\mu}\bar{\psi}^{\db}
+i(\sigma^{\mu}\partial_{\mu}\bar{\psi})_{\alpha}\delta_{\da}{}^{\db}\,,\\
{}\\
{}[\{Q_{\alpha},Q_{\beta}\},\bar{\psi}^{\da}]=0\,,\\
{}\\
{}[\{\bar{Q}_{\da},\bar{Q}_{\db}\},\bar{\psi}^{\dot{\gamma}}]
=i\left(
\delta_{\da}{}^{\dot{\gamma}}(\partial_{\mu}\psi\sigma^{\mu})_{\db}+
\delta_{\db}{}^{\dot{\gamma}}(\partial_{\mu}\psi\sigma^{\mu})_{\da}\right)\,.
\end{array}
\end{equation}
\end{subequations}
\newline

Now it is obvious that  with constraints on $\psi,\,\bar{\psi}$ 
or ``equations of motion''
\begin{equation}
\begin{array}{cc}
\sigma^{\mu}\partial_{\mu}\bar{\psi}=0\,,~~~~&~~~~
\partial_{\mu}\psi\sigma^{\mu}=0\,,
\end{array}
\end{equation}
$4~\!\mbox{D}$ ${\cal N}=1$ on-shell supersymmetric Maxwell theory   admits 
 the following modified supersymmetry algebra\footnote{While this work was being completed, a supersymmetry algebra identical to eq.(\ref{N1susyModi}) but of 
different origin was discussed by Gorsky and Shifman~\cite{9909015}.  
It is relevant to the soliton solutions with the axial geometry - the saturated strings.}
\begin{equation}
\begin{array}{cc}
\multicolumn{2}{c}{\{Q_{\alpha},\bar{Q}_{\da}\}=2\sigma^{\mu}_{\alpha\da}(P_{\mu}+Z_{\mu})\,,}\\
{}&{}\\
{}\{Q_{\alpha},Q_{\beta}\}=0\,,~~~~&~~~~\{\bar{Q}_{\da},\bar{Q}_{\db}\}=0\,,\\
{}&{}\\
{}[P_{\mu},Q_{\alpha}]=0\,,~~~~&~~~~[P_{\mu},\bar{Q}_{\da}]=0\,,\\
{}&{}\\
\multicolumn{2}{c}{[P_{\mu},Z_{\nu}]=0\,,}
\end{array}
\label{N1susyModi}
\end{equation}
where $Z_{\mu}=(Z_{\mu})^{\dagger}$ generates a local gauge transformation
\begin{equation}
\begin{array}{ll}
[Z_{\mu},A_{\nu}]=i\partial_{\nu}A_{\mu}\,,~~~~~&~~~~[Z_{\mu},F_{\nu\rho}]=0\,,\\
{}&{}\\
{}[Z_{\mu},\psi^{\alpha}]=0\,,~~~~&~~~~[Z_{\mu},\bar{\psi}^{\da}]=0\,.
\end{array}
\label{Zgauge}
\end{equation}

Multi-commutators of $Z_{\mu},Q_{\alpha},\bar{Q}_{\da}$  generate infinitely many gauge charges which act on $A_{\mu}$ only and annihilate $\psi^{\alpha},\bar{\psi}^{\da},F_{\mu\nu}$ as in eq.(\ref{Zgauge}). In particular we have
\begin{subeqnarray}
&{}[[Z_{\mu},Q_{\alpha}],A_{\nu}]=-i(\sigma_{\mu}\partial_{\nu}\bar{\psi})_{\alpha}\,,\\
&{}\nonumber\\
&{}[[Z_{\mu},Z_{\nu}],A_{\lambda}]=\partial_{\lambda}F_{\mu\nu}\,,\\
&{}\nonumber\\
&{}[[Z_{\mu_{1}},[Z_{\mu_{2}},\cdots,[Z_{\mu_{n-1}},Z_{\mu_{n}}]\cdots ],A_{\nu}]=
\displaystyle{\prod^{n-2}_{j=1}}(-i\partial_{\mu_{j}})\partial_{\nu}F_{\mu_{n-1}\mu_{n}}\,.
\end{subeqnarray}

\subsection{$4~\!\mbox{D}$ Toy Model of  One-form Superfield : Off-shell\label{YM2}}
Here we study  a $4~\!\mbox{D}$ toy model of a one-form superfield. 
We  assume the supersymmetry
transformation of the lowest order component field is identical to that of the one-form gauge field  in supersymmetric Maxwell theory~(\ref{QAorig}), i.e. 
\begin{equation}
\begin{array}{cc}
[Q_{\alpha},A_{\mu}]=(\sigma_{\mu}\bar{\psi})_{\alpha}\,,~~~~&~~~~
[\bar{Q}_{\da},A_{\mu}]=-(\psi\sigma_{\mu})_{\da}\,,
\end{array}
\label{QA}
\end{equation}
and  require the governing  
supersymmetry algebra to be the standard one consisting of 
$Q_{\alpha},\bar{Q}_{\da},P_{\mu}$ only
\begin{equation}
\begin{array}{cc}
\multicolumn{2}{c}{
\{Q_{\alpha},\bar{Q}_{\da}\}=2\sigma^{\mu}_{\alpha\da}P_{\mu}\,,}\\
{}&{}\\
\{Q_{\alpha},Q_{\beta}\}=0\,,~~~~&~~~~\{\bar{Q}_{\da},\bar{Q}_{\db}\}=0\,.
\end{array}
\label{fsusy}
\end{equation}
We note that eq.(\ref{QA}) is of the most general form without any derivative and gravitinos.  \newline

Provided with  these two equations~(\ref{QA},\ref{fsusy}) alone, without any 
``equation of motion'',    
using some properties of gamma matrices in four-dimensions   
we are going to show that the model is completely solvable at off-shell level,  
possesses a severely constrained kinematics  as 
\begin{itemize}
\item The super multiplet consists of $A_{\mu},\psi^{\alpha},\bar{\psi}^{\da},
\varphi$, where $\varphi$ is a real scalar field.

\item $A_{\mu}$ satisfies the conformal Killing equation
\begin{equation}
\partial_{\mu}A_{\nu}+\partial_{\nu}A_{\mu}=
\textstyle{\frac{1}{2}}(\partial{\cdot A})\,\eta_{\mu\nu}\,,
\label{Killing}
\end{equation}
so that $A_{\mu}$ is at most quadratic in $x$ and 
it has the following  explicit form 
\begin{equation}
A_{\mu}=a_{\mu}+\lambda x_{\mu}+w_{\mu}{}^{\nu}x_{\nu}+2x{\cdot b}\,x_{\mu}-x^{2}b_{\mu}\,,
\label{f12}
\end{equation}
where all the coefficients, $a_{\mu},\lambda,w_{\mu\nu},b_{\mu}$ are Hermitian operators.

\item The supersymmetry transformation rules for the spinor fields, 
$\psi^{\alpha},\bar{\psi}^{\da}$, are
\begin{equation}
\begin{array}{cc}
\multicolumn{2}{c}{\{Q_{\alpha},\psi^{\beta}\}=i\textstyle{\frac{1}{4}}
(\sigma^{[\mu}\tilde{\sigma}^{\nu]})_{\alpha}{}^{\beta}F_{\mu\nu}+
(\varphi+i\textstyle{\frac{1}{4}}\partial{\cdot A})\delta_{\alpha}{}^{\beta}\,,}\\
{}&{}\\
\multicolumn{2}{c}{\{\bar{Q}_{\da},\bar{\psi}^{\db}\}
=i\textstyle{\frac{1}{4}}(\tilde{\sigma}^{[\mu}\sigma^{\nu]})^{\db}{}_{\da}F_{\mu\nu}+
(\varphi-i\textstyle{\frac{1}{4}}\partial{\cdot A})\delta^{\db}{}_{\da}\,,
}\\
{}&{}\\
\{Q_{\alpha},\bar{\psi}^{\da}\}=0\,,~~~~&~~~~\{\bar{Q}_{\da},\psi^{\alpha}\}=0\,.
\end{array}
\label{f32}
\end{equation}

\item $\psi^{\alpha},\bar{\psi}^{\da}$ are at most linear in $x$. The explicit forms are
\begin{equation}
\begin{array}{cc}
\psi^{\alpha}=\xi^{\alpha}+(\bar{\rho}\tx)^{\alpha}\,,~~~~&~~~~
\bar{\psi}^{\da}=\bar{\xi}^{\da}+(\tx\rho)^{\da}\,,
\end{array}
\label{fpsiform} 
\end{equation}
where $\tx=x_{\mu}\tilde{\sigma}^{\mu}$ and  
the lower spinorial indices of $\rho_{\alpha},\bar{\rho}_{\da}$ are to be 
understood.                                                                                                                                                                                                                                                                                                                                                                                                                                                                                                                                                                                                                                                                                                                     

\item The real scalar field, $\varphi$, is a spacetime independent constant 
field.    Its supersymmetry transformation rules are
\begin{equation}
\begin{array}{cc}
[Q_{\alpha},\varphi]=-i\textstyle{\frac{3}{4}}(\sigma^{\mu}\partial_{\mu}\bar{\psi})_{\alpha}\,,~~~~&~~~~
[\bar{Q}_{\da},\varphi]=-i\textstyle{\frac{3}{4}}(\partial_{\mu}\psi\sigma^{\mu})_{\da}\,.
\label{Qchi}
\end{array}
\end{equation}

\item The supersymmetry algebra~(\ref{fsusy}) is represented by
\begin{subequations}
\label{rigidYM}
\begin{equation}
\begin{array}{ll}
[Q_{\alpha},a_{\mu}]=(\sigma_{\mu}\bar{\xi})_{\alpha}\,,~~~~&~~~~
[\bar{Q}_{\da},a_{\mu}]=-(\xi\sigma_{\mu})_{\da}\,,\\
{}&{}\\
{}[Q_{\alpha},\lambda]=\rho_{\alpha}\,,~~~~&~~~~
[\bar{Q}_{\da},\lambda]=-\bar{\rho}_{\da}\,,\\
{}&{}\\
{}[Q_{\alpha},w_{\mu\nu}]=(\sigma_{[\mu}\tilde{\sigma}_{\nu]}\rho)_{\alpha}
\,,~~~~&~~~~
[\bar{Q}_{\da},w_{\mu\nu}]=(\bar{\rho}\tilde{\sigma}_{[\mu}\sigma_{\nu]})_{\da}
\,,\\
{}&{}\\
{}[Q_{\alpha},b_{\mu}]=0\,,~~~~&~~~~
[\bar{Q}_{\da},b_{\mu}]=0\,,
\end{array}
\label{f18}
\end{equation}
\newline
\begin{equation}
\begin{array}{l}
[P_{\mu},a_{\nu}]=i(w_{\mu\nu}-\lambda\eta_{\mu\nu})\,,\\
{}\\
{}[P_{\mu},\lambda]=-2ib_{\mu}\,,\\
{}\\
{}[P_{\lambda},w_{\mu\nu}]=2i(b_{\mu}\eta_{\nu\lambda}-b_{\nu}\eta_{\mu\lambda})\,,\\
{}\\
{}[P_{\mu},b_{\nu}]=0\,,
\end{array}
\label{f19}
\end{equation}
\newline
\begin{equation}
\begin{array}{ll}
\{Q_{\alpha},\bar{\xi}^{\da}\}=0\,,~~~~~~~~~~~&~~~~
\{\bar{Q}_{\da},\xi^{\alpha}\}=0\,,~~~~~~~~~~~\\
{}&{}\\
\{Q_{\alpha},\rho_{\beta}\}=0\,,~~~~~~~~~~~&~~~~
\{\bar{Q}_{\da},\bar{\rho}_{\db}\}=0\,,~~~~~~~~~~~
\end{array}
\label{f21}
\end{equation}
\newline
\begin{equation}
\begin{array}{l}
\{Q_{\alpha},\xi^{\beta}\}=-i\textstyle{\frac{1}{2}}w_{\mu\nu}(\sigma^{[\mu}\tilde{\sigma}^{\nu]})_{\alpha}{}^{\beta}
+(\varphi+i\lambda)\delta_{\alpha}{}^{\beta}\,,\\
{}\\
\{\bar{Q}_{\da},\bar{\xi}^{\db}\}=-i\textstyle{\frac{1}{2}}w_{\mu\nu}(\tilde{\sigma}^{[\mu}\sigma^{\nu]})^{\db}{}_{\da}
+(\varphi-i\lambda)\delta^{\db}{}_{\da}\,,
\end{array}
\label{f26}
\end{equation}
where
\begin{equation}
\varphi=\textstyle{\frac{1}{4}}(\{Q_{\alpha},\xi^{\alpha}\}+\{\bar{Q}_{\da},\bar{\xi}^{\da}\})=\varphi^{\dagger}\,,
\label{f27}
\end{equation}
\newline
\begin{equation}
\begin{array}{ll}
{}\{Q_{\alpha},\bar{\rho}_{\da}\}=2i(b{\cdot\sigma})_{\alpha\da}\,,~~~~&~~~~
\{\bar{Q}_{\da},\rho_{\alpha}\}=-2i(b{\cdot\sigma})_{\alpha\da}\,,
\end{array}
\label{f23}
\end{equation}
\newline
\begin{equation}
\begin{array}{ll}
{}[P_{\mu},\xi^{\alpha}]
=-i(\bar{\rho}\tilde{\sigma}_{\mu})^{\alpha}\,,~~~~&~~~~
[P_{\mu},\bar{\xi}^{\da}]=-i(\tilde{\sigma}_{\mu}\rho)^{\da}\,,\\
{}&{}\\
{}[P_{\mu},\rho_{\alpha}]=0\,,~~~~&~~~~[P_{\mu},\bar{\rho}_{\da}]=0\,,
\end{array}
\label{f20}
\end{equation}
\newline
\begin{equation}
\begin{array}{ll}
[Q_{\alpha},\varphi]=-3i\rho_{\alpha}\,,~~~~&~~~~
[\bar{Q}_{\da},\varphi]=-3i\bar{\rho}_{\da}\,,\\
{}&{}\\
\multicolumn{2}{c}{[P_{\mu},\varphi]=0\,.}
\end{array}
\label{Qchirho}
\end{equation}
\end{subequations}
\end{itemize}

{}~\newline
\indent\textit{Proof}\newline
\newline
A Jacobi identity analogous 
to eq.(\ref{preJacobi}) gives using eq.(\ref{fcontract})
\begin{equation}
\partial_{\mu}A_{\nu}=i\textstyle{\frac{1}{4}}\left(
(\tilde{\sigma}_{\mu}\sigma_{\nu})^{\da}{}_{\db}\{\bar{Q}_{\da},
\bar{\psi}^{\db}\}-({\sigma}_{\nu}\tilde{\sigma}_{\mu})_{\beta}{}^{\alpha}
\{Q_{\alpha},\psi^{\beta}\}\right)\,,
\label{f1}
\end{equation}
while Jacobi identity for $\{Q_{\alpha},Q_{\beta}\}=0$ gives with 
$A^{\da\alpha}\equiv A_{\mu}\tilde{\sigma}^{\mu\da\alpha}$
\begin{equation}
\begin{array}{ll}
0&=\{Q_{\alpha},[Q_{\beta},A^{\dot{\gamma}\gamma}]\}+
\{Q_{\beta},[Q_{\alpha},A^{\dot{\gamma}\gamma}]\}\\
{}&{}\\
{}&=2\delta_{\alpha}{}^{\gamma}\{Q_{\beta},\bar{\psi}^{\dot{\gamma}}\}+
2\delta_{\beta}{}^{\gamma}\{Q_{\alpha},\bar{\psi}^{\dot{\gamma}}\}\,,
\end{array}
\end{equation}
and hence
\begin{equation}
\begin{array}{cc}
\{Q_{\alpha},\bar{\psi}^{\da}\}=0\,,~~~~&~~~~
\{\bar{Q}_{\da},\psi^{\alpha}\}=0\,.
\end{array}
\label{f1/2}
\end{equation}
Eq.(\ref{f1}) is equivalent to
\begin{subeqnarray}
\partial_{\mu}A_{\nu}+\partial_{\nu}A_{\mu}=i\textstyle{\frac{1}{2}}\left(
\{\bar{Q}_{\da},\bar{\psi}^{\da}\}-\{Q_{\alpha},\psi^{\alpha}\}\right)
\,\eta_{\mu\nu}\,,~~~~~~~~\label{f2}\\
\nonumber\\
F_{\mu\nu}=i\textstyle{\frac{1}{2}}\left(
(\tilde{\sigma}_{[\mu}\sigma_{\nu]})^{\da}{}_{\db}\{\bar{Q}_{\da},
\bar{\psi}^{\db}\}+({\sigma}_{[\mu}\tilde{\sigma}_{\nu]})_{\beta}{}^{\alpha}
\{Q_{\alpha},\psi^{\beta}\}\right)\,.\label{f3}
\end{subeqnarray}
Eq.(\ref{f2}) gives the conformal Killing equation~(\ref{Killing}) with the general solution~(\ref{f12}) in four-dimensions~\cite{ginsparg} and 
eq.(\ref{f3}) gives from eq.(\ref{fsig4}) 
\begin{equation}
\{Q_{\alpha},\psi^{\beta}\}-\textstyle{\frac{1}{2}}\delta_{\alpha}{}^{\beta}
\{Q_{\gamma},\psi^{\gamma}\}=i\textstyle{\frac{1}{4}}F_{\mu\nu}
({\sigma}^{[\mu}\tilde{\sigma}^{\nu]})_{\alpha}{}^{\beta}\,,
\label{f4}
\end{equation}
so that from eqs.(\ref{fsusy},\ref{f1/2}) 
\begin{equation}
\begin{array}{ll}
i\textstyle{\frac{1}{4}}
({\sigma}^{[\mu}\tilde{\sigma}^{\nu]})_{\alpha}{}^{\beta}
[\bar{Q}_{\da},F_{\mu\nu}]  
& =-i\textstyle{\frac{1}{2}}
({\sigma}^{[\mu}\tilde{\sigma}^{\nu]})_{\alpha}{}^{\beta}
(\partial_{\mu}\psi\sigma_{\nu})_{\da}\\
{}&{}\\
{}&=[\bar{Q}_{\da},\{Q_{\alpha},\psi^{\beta}\}]-
\textstyle{\frac{1}{2}}\delta_{\alpha}{}^{\beta}
[\bar{Q}_{\da},\{Q_{\gamma},\psi^{\gamma}\}]\\
{}&{}\\
{}&=-2i\sigma^{\mu}_{\alpha\da}\partial_{\mu}\psi^{\beta}+
i\delta_{\alpha}{}^{\beta}(\partial_{\mu}\psi\sigma^{\mu})_{\da}\,.
\end{array}
\label{QQpsi}
\end{equation}
Using eqs.(\ref{QA},\ref{sigti},\ref{fese},\ref{fsig4}) one can show that 
eq.(\ref{QQpsi}) is equivalent to the following simple formula
\begin{equation}
\partial_{\mu}\psi^{\alpha}=\textstyle{\frac{1}{4}}(\partial_{\nu}\psi\sigma^{\nu}\tilde{\sigma}_{\mu})^{\alpha}\,.
\label{f7}
\end{equation}
This gives
\begin{equation}
\begin{array}{ll}
\partial_{\mu}\partial_{\nu}\psi&=(\textstyle{\frac{1}{4}})^{2}
\partial_{\rho}\partial_{\lambda}\psi\sigma^{\lambda}\tilde{\sigma}_{\mu}\sigma^{\rho}\tilde{\sigma}_{\nu}\\
{}&{}\\
{}&=(\textstyle{\frac{1}{4}})^{2}
\partial_{\rho}\partial_{\lambda}\psi\sigma^{\lambda}(2\delta_{\mu}{}^{\rho}-
\tilde{\sigma}^{\rho}\sigma_{\mu})\tilde{\sigma}_{\nu}\\
{}&{}\\
{}&=\textstyle{\frac{1}{2}}\partial_{\mu}\partial_{\nu}\psi-
\textstyle{\frac{1}{16}}\Box\psi\sigma_{\mu}\tilde{\sigma}_{\nu}\,,
\end{array}
\end{equation}
so that
\begin{equation}
\begin{array}{l}
\Box\psi^{\alpha}=0\,,\\
{}\\
\partial_{\mu}\partial_{\nu}\psi^{\alpha}=0\,.
\end{array}
\end{equation}
Thus $\psi^{\alpha},\bar{\psi}^{\da}$ are at most linear in spacetime coordinates, $x$, and from eq.(\ref{f7}) we can write their general forms as in eq.(\ref{fpsiform}). \newline

Now from eqs.(\ref{QA},\ref{f12},\ref{f1/2}) we get eq.(\ref{f18},\ref{f21}),  
and from $[P_{\mu},A_{\nu}]=-i\partial_{\mu}A_{\nu},$  
$[P_{\mu},\psi]=-i\partial_{\mu}\psi$   we get eqs.(\ref{f19},\ref{f20}). 
Substituting eqs.(\ref{f12},\ref{fpsiform}) into eq.(\ref{f4}) gives
\begin{subeqnarray}
\{Q_{\alpha},\xi^{\beta}\}-\textstyle{\frac{1}{2}}\delta_{\alpha}{}^{\beta}
\{Q_{\gamma},\xi^{\gamma}\}=-i\textstyle{\frac{1}{2}}w_{\mu\nu}
({\sigma}^{[\mu}\tilde{\sigma}^{\nu]})_{\alpha}{}^{\beta}\,,~~~~~~~~~~~~
\label{f22}\\
{}\nonumber\\
\{Q_{\alpha},(\rho\tilde{\sigma}_{\mu})^{\beta}\}-\textstyle{\frac{1}{2}}\delta_{\alpha}{}^{\beta}\{Q_{\gamma},(\rho\tilde{\sigma}_{\mu})^{\gamma}\}=
i(b{\cdot\sigma}\tilde{\sigma}_{\mu}
-\sigma_{\mu}b{\cdot\tilde{\sigma}})_{\alpha}{}^{\beta}\,.
\label{latterQrho}
\end{subeqnarray}
Contracting the latter with $\sigma^{\mu}_{\beta\da}$ gives eq.(\ref{f23}) 
which is however equivalent to eq.(\ref{latterQrho}). \newline
Eq.(\ref{f2}) gives 
\begin{equation}
\{Q_{\alpha},\xi^{\alpha}\}-\{\bar{Q}_{\da},\bar{\xi}^{\da}\}=4i\lambda\,.
\label{f24}
\end{equation}
Now we define $\varphi$ as in eq.(\ref{f27}). Then from eqs.(\ref{f22},\ref{f24}) we get eq.(\ref{f26}). The supersymmetry transformation rules for $\varphi$~(\ref{Qchirho}) follow from eqs.(\ref{f18},\ref{f21},\ref{f26},\ref{f27},\ref{f20}), since
\begin{equation}
\begin{array}{ll}
[Q_{\alpha},\varphi]&=\textstyle{\frac{1}{2}}[Q_{\alpha},\varphi+i\lambda]+
\textstyle{\frac{1}{2}}\sigma^{\mu}_{\alpha\da}[P_{\mu},\bar{\xi}^{\da}]\\
{}&{}\\
{}&=\textstyle{\frac{1}{2}}[Q_{\alpha},\varphi]
-i\textstyle{\frac{3}{2}}\rho_{\alpha}\,.
\end{array}
\end{equation}
~~~{}\newline

Finally one can check that eq.(\ref{rigidYM}) is consistent with the supersymmetry algebra~(\ref{fsusy}).\newline

\begin{flushright}\textit{Q.E.D.}\end{flushright}

\newpage

\section{Summary and Discussion}
In section \ref{6DTENSOR},  we reconstructed  $6~\!\mbox{D}$ 
$(1,0)$ tensor multiplet at off-shell and on-shell level respectively: At off-shell level, 
within a superfield formalism in an algebraic way, we showed that   
the existence of a two-form gauge field follows only implicitly as 
the exterior derivative of the three-form tensor vanishes. 
We also exhibited  an explicit expression for the corresponding 
superfield.  At on-shell level,  we analysed the supersymmetry algebra  
when the supersymmetry transformation rule 
for the   two-form gauge field is given explicitly.  We proposed   
a notion for  \textit{gauge charges} carrying    spacetime/spinor indices.  
Along with translations  and  super charges they form an infinite dimensional super algebra formally.    
This has to do with the fact that  there exist  infinite degrees of freedom for local   
gauge transformations. \newline

In section \ref{TENSOR2}, we investigated the consequences of  
``switching off''  the gauge charges  by considering a 
$6~\!\mbox{D}$ off-shell toy model of a two-form superfield .  
We made two assumptions there. One is that the  
supersymmetry transformation of  the  lowest component field of the superfield, $B_{AB}$,  is identical to the that of the two-form gauge field in on-shell $(1,0)$  tensor multiplet and   
the other is that the model is subject to 
the standard six-dimensional $(N,0)$ supersymmetry algebra
 which consists of  $P,Q$ only with an arbitrary natural number, $N$. 
With these two assumptions alone, we showed 
that the model is completely solvable at off-shell level and possesses a severely constrained kinematics in the sense that: (i) $B_{AB},\,\psi^{i}_{\alpha}$ 
alone form a super multiplet without introducing any scalar field, 
(ii) The supersymmetry transformation rule for $\psi^{i}$ is   
determined,  (iii) The field strength of the two-form field is automatically self-dual,  (iv) $B_{AB}$ is at most linear in spacetime coordinates, $x$, and 
$\psi^{i}$ 
is a spacetime independent constant field, (v) It is possible to determine how the coefficients  in the expansion of each component field in $x$ transform to another under supersymmetry transformations, i.e. these coefficients form a ``rigid'' representation of the supersymmetry algebra.\newline

In section \ref{YM}, similar analysis on $4~\!\mbox{D}$  
${\cal N}=1$ supersymmetric Maxwell theory was performed and we  identified an infinite dimensional supersymmetry algebra with gauge charges carrying  
spacetime/spinor indices. We also analysed  
a $4~\!\mbox{D}$ toy model of a one-form superfield.  We found  
the model is completely solvable at off-shell level too as   (i) The super multiplet consists of 
$A_{\mu},\psi^{\alpha},\bar{\psi}^{\da}$ and a real scalar field $\varphi$, 
(ii) The supersymmetry transformation rules for all the component fields are 
determined,  (iii) $A_{\mu}$ satisfies the conformal Killing equation 
so that $A_{\mu}$ is at most quadratic in $x$,  
(iv)$\psi^{\alpha},\bar{\psi}^{\da}$ are at most linear in $x$, and   
$\varphi$ is a spacetime independent constant field, 
(v) The coefficients  in the expansion of each component field in $x$ 
form  a ``rigid'' representation of the supersymmetry algebra.\newline

Our study of the toy model signifies the importance of 
the gauge symmetry/charges in ordinary gauge theories and reveals the power of supersymmetry once more.

\newpage

\appendix
\begin{center}
\Large{\textbf{Appendix}}
\end{center}

Here we exhibit some useful identities on  gamma matrices in six and four 
dimensions  which were previously given in Ref. \cite{paper2} with  details. 
\section{Gamma Matrices in Six-dimensions}
With the Minkowskian metric $\eta^{AB}=\mbox{diag}(+1,-1,-1,\cdots,-1)$, 
the $4\times 4$ matrices, $\gamma^{A},\tilde{\gamma}^{A}$
satisfy 
\begin{equation}
\gamma^{A}\tilde{\gamma}^{B}+\gamma^{B}\tilde{\gamma}^{A}=2\eta^{AB}\,,
\label{algebra}
\end{equation}
and hence\footnote{We put $\epsilon^{012345}=1$.} 
\begin{eqnarray}
&\mbox{tr}(\gamma^{A}\tilde{\gamma}^{B})=4\eta^{AB}\,,&\label{tr2}\\
{}\nonumber\\
&\mbox{tr}(\gamma^{A}\tilde{\gamma}^{B}\gamma^{C}\tilde{\gamma}^{D})
=4(\eta^{AB}\eta^{CD}-\eta^{AC}\eta^{BD}+\eta^{BC}\eta^{DA})\,,&\label{tr4}\\
{}\nonumber\\
&\mbox{tr}(\gamma^{[A}\tilde{\gamma}^{B}\gamma^{C]}
\tilde{\gamma}_{[D}\gamma_{E}\tilde{\gamma}_{F]})=4\epsilon^{ABC}{}_{DEF}
-24\delta^{[A}_{~D}\delta^{B}_{~E}\delta^{C]}_{~F}\,,&\label{tr6}\\
{}\nonumber\\
&\gamma^{0}\gamma^{A\dagger}\gamma^{0}=\gamma^{A}\,,~~~~~~~~~
\gamma^{A\dagger}=\tilde{\gamma}_{A}\,.\label{hermiticity}&
\end{eqnarray}
\newline

In six-dimensions, without loss of generality, 
$\gamma^{A},\,\tilde{\gamma}^{A}$ may be taken 
to be anti-symmetric
\begin{equation}
\begin{array}{cc}
(\gamma^{A})_{\alpha\beta}=-(\gamma^{A})_{\beta\alpha}\,,~~~~ &~~~~ 
(\tilde{\gamma}^{A})^{\alpha\beta}=-(\tilde{\gamma}^{A})^{\beta\alpha}\,,
\end{array}
\label{gamma}
\end{equation}
and hence 
$\{\gamma^{A}\}$ and $\{\tilde{\gamma}^{A}\}$ are  separately bases 
of $4 \times 4$ anti-symmetric matrices so that
\begin{equation}
(\gamma^{A})_{\alpha\beta}(\tilde{\gamma}_{A})^{\gamma\delta}=
2(\delta^{~\delta}_{\alpha}\delta^{~\gamma}_{\beta}-
\delta^{~\gamma}_{\alpha}\delta^{~\delta}_{\beta})\,. 
\label{anti-sym}
\end{equation}
\newline

Due to  the identities
\begin{equation}
\begin{array}{cc}
\gamma^{[A}\tilde{\gamma}^{B}\gamma^{C]}=-\textstyle{\frac{1}{6}}
\epsilon^{ABC}{}_{DEF}\,\gamma^{D}\tilde{\gamma}^{E}\gamma^{F}\,,~~~&~~~
\tilde{\gamma}^{[A}\gamma^{B}\tilde{\gamma}^{C]}=\textstyle{\frac{1}{6}}
\epsilon^{ABC}{}_{DEF}\,\tilde{\gamma}^{D}\gamma^{E}\tilde{\gamma}^{F}\,,
\end{array}
\label{Due}
\end{equation}
there are only $10$ independent $\gamma^{[A}\tilde{\gamma}^{B}\gamma^{C]}$ and 
$\tilde{\gamma}^{[A}\gamma^{B}\tilde{\gamma}^{C]}$  separately and 
 both of them form   bases 
of $4 \times 4$ symmetric matrices  with the completeness relation 
\begin{equation}
(\gamma^{[A}\tilde{\gamma}^{B}\gamma^{C]})_{\alpha\beta}
(\tilde{\gamma}_{[A}\gamma_{B}\tilde{\gamma}_{C]})^{\gamma\delta}=
-24(\delta^{~\gamma}_{\alpha}\delta^{~\delta}_{\beta}
+\delta^{~\delta}_{\alpha}\delta^{~\gamma}_{\beta})\,.
\label{sym} 
\end{equation}
The coefficient on the right hand side may be determined by eq.(\ref{tr6}). \newline

$\{\gamma^{[A}\tilde{\gamma}^{B]}, 1\}$ also forms  
a basis of general $4\times 4$ matrices with the completeness relation
\begin{equation}
-\textstyle{\frac{1}{8}}(\gamma^{[A}\tilde{\gamma}^{B]})_{\alpha}^{~\beta}
(\gamma_{[A}\tilde{\gamma}_{B]})_{\gamma}^{~\delta}+
\textstyle{\frac{1}{4}}\delta_{\alpha}^{~\beta}\delta_{\gamma}^{~\delta}=
\delta_{\alpha}^{~\delta}\delta_{\gamma}^{~\beta}\,.
\label{complete}
\end{equation}
\newline

With this choice of gamma matrices we get
\begin{subeqnarray}
\label{LEQ}
&\gamma^{A}_{\alpha\beta}=\textstyle{\frac{1}{2}}
\epsilon_{\alpha\beta\gamma\delta}\tilde{\gamma}^{A}{}^{\gamma\delta}\,,
\,\,\,\,\,\,\,\,\,\,\,\,
\tilde{\gamma}^{A}{}^{\alpha\beta}=\textstyle{\frac{1}{2}}
\epsilon^{\alpha\beta\gamma\delta}\gamma^{A}_{\gamma\delta}\,,&\label{gtg}\\
{}\nonumber\\
&\gamma^{[A}\tilde{\gamma}^{B]}\gamma^{C}
-(\gamma^{[A}\tilde{\gamma}^{B]}\gamma^{C})^{t}=2\eta^{BC}\gamma^{A}
-2\eta^{CA}\gamma^{B}\,,&\label{minus}\\
{}\nonumber\\
&\gamma^{[A}\tilde{\gamma}^{B]}\gamma^{C}
+(\gamma^{[A}\tilde{\gamma}^{B]}\gamma^{C})^{t}=
2\gamma^{[A}\tilde{\gamma}^{B}\gamma^{C]}\,,&\label{plus}\\
{}\nonumber\\
&\gamma^{[A}\tilde{\gamma}^{B]}\gamma_{[C}\tilde{\gamma}_{D}\gamma_{E]}
-(\gamma^{[A}\tilde{\gamma}^{B]}\gamma_{[C}\tilde{\gamma}_{D}\gamma_{E]})^{t}
=2\epsilon^{AB}{}_{CDEF}\gamma^{F}
-12\delta^{A}{}_{[C}\delta^{B}{}_{D}\gamma_{E]}\,,&\label{minus2}\\
{}\nonumber\\
&\gamma^{A}_{\alpha\beta}(\gamma_{A}\tilde{\gamma}^{B})_{\gamma}{}^{\delta}=
2\delta_{\alpha}{}^{\delta}\gamma^{B}_{\beta\gamma}+2\delta_{\beta}{}^{\delta}\gamma^{B}_{\gamma\alpha}
+2\delta_{\gamma}{}^{\delta}\gamma^{B}_{\alpha\beta}\,,&\label{L8}\\
{}\nonumber\\
&(\gamma^{[A}\tilde{\gamma}^{B]})_{\alpha}{}^{\beta}\gamma_{B\gamma\delta}=
2\delta^{\beta}{}_{\gamma}\gamma^{A}_{\alpha\delta}-
2\delta^{\beta}{}_{\delta}\gamma^{A}_{\alpha\gamma}-
\delta_{\alpha}{}^{\beta}\gamma^{A}_{\gamma\delta}\,,&\label{L6}\\
{}\nonumber\\
&(\gamma^{[A}\tilde{\gamma}^{B}\gamma^{C]})_{\alpha\beta}
(\gamma_{[A}\tilde{\gamma}_{B]})_{\gamma}{}^{\delta}=
4(\delta_{\alpha}{}^{\delta}\gamma^{C}_{\beta\gamma}+
\delta_{\beta}{}^{\delta}\gamma^{C}_{\alpha\gamma})\,,&\label{L7}\\
{}\nonumber\\
&\gamma_{[A}\tilde{\gamma}_{B}\gamma_{C]}\tilde{\gamma}^{D}=
3\gamma_{[A}\tilde{\gamma}_{B}\delta_{C]}{}^{D}-\textstyle{\frac{1}{2}}
\epsilon_{ABC}{}^{DEF}\gamma_{E}\tilde{\gamma}_{F}\,,&\\
{}\nonumber\\
&\gamma^{D}\tilde{\gamma}_{[A}\gamma_{B}\tilde{\gamma}_{C]}=
3\gamma_{[A}\tilde{\gamma}_{B}\delta_{C]}{}^{D}+\textstyle{\frac{1}{2}}
\epsilon_{ABC}{}^{DEF}\gamma_{E}\tilde{\gamma}_{F}\,,&\\
{}\nonumber\\
&6\gamma_{[A}\tilde{\gamma}_{B}\delta_{C]}{}^{D}=\gamma_{[A}\tilde{\gamma}_{B}\gamma_{C]}\tilde{\gamma}^{D}+\gamma^{D}\tilde{\gamma}_{[A}\gamma_{B}\tilde{\gamma}_{C]}\,.&\label{6gamma}
\end{subeqnarray}
It is useful to note
\begin{equation}
\epsilon_{ABCDIJ}\epsilon^{EFGHIJ}=-48\delta_{A}^{~[E}\delta_{B}^{~F}
\delta_{C}^{~G}\delta_{D}^{~H]}\,.
\label{epep6}
\end{equation}   
\section{Sigma Matrices in Four-dimensions}
The $2\times 2$ matrices, $\sigma^{\mu},\tilde{\sigma}^{\mu}$
satisfy
\begin{equation}
\sigma^{\mu}\tilde{\sigma}^{\nu}+\sigma^{\nu}\tilde{\sigma}^{\mu}
=2\eta^{\mu\nu}\,,
\label{sigti}
\end{equation}
and hence
\begin{subeqnarray}
&\textstyle{\frac{1}{2}}\mbox{tr}(\sigma^{\mu}\tilde{\sigma}^{\nu})
=\eta^{\mu\nu}\,,&\label{fTr2}\\
{}\nonumber\\
&\textstyle{\frac{1}{2}}\mbox{tr}(\sigma^{\mu}\tilde{\sigma}^{\nu}
\sigma^{\lambda}\tilde{\sigma}^{\rho})=\eta^{\mu\nu}\eta^{\lambda\rho}
+\eta^{\nu\lambda}\eta^{\rho\mu}-\eta^{\mu\lambda}\eta^{\nu\rho}-
i\epsilon^{\mu\nu\lambda\rho}\,,&\label{fTr4}
\end{subeqnarray}
where we put $\epsilon_{0123}=-\epsilon^{0123}=1$.\newline

$\sigma^{\mu}$ and $\tilde{\sigma}^{\mu}$  separately form bases 
of $2 \times 2$  matrices with the completeness relation
\begin{equation}
\sigma^{\mu}_{\alpha\dot{\alpha}}\tilde{\sigma}_{\mu}^{\dot{\beta}\beta}=
2\delta_{\alpha}^{~\beta}\delta_{\dot{\alpha}}^{~\dot{\beta}}\,.
\label{fcontract}
\end{equation}
The coefficient on the right hand side 
may be determined by eq.(\ref{fTr2}). \newline

$\sigma^{\mu}$ and $\tilde{\sigma}^{\mu}$ are related by
\begin{equation}
\epsilon\tilde{\sigma}^{\mu t}\bar{\epsilon}=-\sigma^{\mu}\,.
\label{fese}
\end{equation}
where
$\epsilon_{\alpha\beta},\,\bar{\epsilon}_{\dot{\alpha}\dot{\beta}}$ are 
$2\times 2$ anti-symmetric matrices, 
$\epsilon_{12}=\bar{\epsilon}_{12}=1$ with inverses,\newline 
$(\epsilon^{-1})^{\alpha\beta},\,
(\bar{\epsilon}^{-1})^{\dot{\alpha}\dot{\beta}}$.\newline

From eqs.(\ref{fcontract},\,\ref{fese}) we get
\begin{equation}
\begin{array}{c}
\sigma^{\mu}_{\alpha\da}\sigma_{\mu\beta\db}=2\epsilon_{\alpha\beta}
\bar{\epsilon}_{\da\db}\,,\\
{}\\
(\sigma^{[\mu}\tilde{\sigma}^{\nu ]})_{\alpha}{}^{\beta}
(\sigma_{[\mu}\tilde{\sigma}_{\nu ]})_{\gamma}{}^{\delta}=
4(\delta_{\alpha}^{~\beta}\delta_{\gamma}^{~\delta}
-2\delta_{\alpha}^{~\delta}\delta^{~\beta}_{\gamma})\,,\\
{}\\
(\sigma^{[\mu}\tilde{\sigma}^{\nu ]})_{\alpha}{}^{\beta}
(\tilde{\sigma}_{[\mu}\sigma_{\nu ]})^{\da}{}_{\db}=0\,.
\label{fsig4}
\end{array}
\end{equation}
It is useful to note
\begin{equation}
\delta_{\alpha}^{~\delta}\delta^{~\beta}_{\gamma}-
\delta_{\alpha}^{~\beta}\delta^{~\delta}_{\gamma}
=\epsilon_{\alpha\gamma}\epsilon^{-1}{}^{\beta\delta}\,.
\label{fdde}
\end{equation}


\newpage
\bibliographystyle{unsrt}
\bibliography{reference}

\end{document}